\documentclass[a4paper,11pt]{article}
\pdfoutput=1 

\usepackage{jheppub} 

\usepackage[T1]{fontenc} 

\usepackage{amssymb} 
\usepackage{amsmath} %
\usepackage{color}
\usepackage{booktabs} 
\usepackage{array} 
\usepackage{paralist} 
\usepackage{verbatim} 
\usepackage{subfig} 
\usepackage{placeins,tikz}
\usepackage{slashed}

\usepackage{hyperref}

\newcommand{\be}{\begin{equation}}
\newcommand{\ee}{\end{equation}}

\newcommand{\R}{\mathbb{R}}

\newcommand{\N}{\mathcal{N}}

\date{} 

\begin{document}
\begin{titlepage}

\begin{center}
 {\LARGE\bfseries  A double copy for ${\mathcal N}=2$ supergravity:  \\    \vskip 2mm
 a linearised tale told on-shell }
\\[10mm]

\textbf{G.L.~Cardoso, S. Nagy and S.~Nampuri}

\vskip 6mm
{\em  Center for Mathematical Analysis, Geometry and Dynamical Systems,\\
  Department of Mathematics, 
  Instituto Superior T\'ecnico,\\ Universidade de Lisboa,
  Av. Rovisco Pais, 1049-001 Lisboa, Portugal}\\[10mm]

{\tt gcardoso@math.tecnico.ulisboa.pt}\;,\;\,{\tt  snagy@math.tecnico.ulisboa.pt}\;,\;\,{\tt
  nampuri@gmail.com}
\end{center}

\vskip .2in
\begin{center} {\bf ABSTRACT } \end{center}
\begin{quotation}\noindent 
We construct the on-shell double copy dictionary for linearised four-dimensional ${\mathcal N}=2$ supergravity coupled to one vector multiplet
with a quadratic prepotential. We apply this dictionary to the weak-field approximation of dyonic BPS black holes in this theory.

\end{quotation}
\vfill
\today
\end{titlepage}

\tableofcontents

\section{Introduction}
One of the richest and most outstanding pursuits in gravitation for the past decades
has been the attempt to formulate gravity in terms of gauge field degrees of freedom. One approach towards this goal is the gauge-gravity duality approach which seeks to describe gravitational degrees of freedom as holographically  encoded in terms of a lower dimensional field theory. The AdS/CFT correspondence 
\cite{Maldacena:1997re,Gubser:1998bc,Witten:1998qj}, 
which equates the gravitational path integral in the bulk with given boundary conditions for the fields to the path integral in a lower dimensional CFT, is the pinnacle of achievement based on this theme. A more kinematically flavored approach is based on rewriting gravity amplitudes as double copies of gauge theory amplitudes \cite{Bianchi:2008pu,Bern:2008qj,Bern:2009kd,Bern:2010ue,Bern:2010yg,Huang:2012wr,
Carrasco:2012ca,Chiodaroli:2014xia,Chiodaroli:2016jqw}. 
The double copy approach\footnote{See \cite{Borsten:2015pla} for a review.} indicates that heuristically, to rewrite gravity in terms of squared gauge theories, one must replace gravitational fields by a tensor product of appropriate gauge fields terms \cite{Siegel:1988qu,Siegel:1995px,Borsten:2013bp,Anastasiou:2013hba, Anastasiou:2014qba,Anastasiou:2015vba,Anastasiou:2016xxx}. The holographic approach has yielded invaluable insights into the non-perturbative structure of gravity in terms of the organization of dual CFT data, but it applies to spacetimes that are asymptotically
AdS. The second approach
yields the rewriting of gravitational amplitudes in flat spacetime, in terms of gauge theory amplitudes. However, it
has not yet produced direct knowledge of solitonic configurations, as computing amplitudes in non-trivial backgrounds remains a mathematically challenging exercise, to date. Motivated by these considerations, here, we adopt the 'Gravity as a double copy of gauge theories' philosophy inspired by the second approach and initiate a program to develop a prescription for mapping on-shell configurations of $D=4, \mathcal{N}=2$ ungauged supergravity theories to a double copy description. For the purposes of this note, which is to demonstrate the existence of such a consistent  lexicon, we restrict ourselves to one of the simplest ${\mathcal N}=2$ supergravity theories,
describing the coupling of one vector multiplet to supergravity, namely the one based
on a quadratic prepotential given by $F=- i X^0 X^1$.

The first step here is to verify a match between the on-shell degrees of freedom in gauge and gravity theories. The next step is to 
propose a mathematical structure that allows gravitational fields to be written as tensor-like combinations of the gauge fields,  and which naturally incorporate a map between symmetries on the gravity side, such as diffeomorphism, to local gauge symmetries in the double copy description \cite{Anastasiou:2014qba}. At the linearised approximation level of gravity, inspired by the results of the amplitude calculations which indicate that fields like the graviton should be replaced by a tensor product of gauge fields in momentum space, \cite{Anastasiou:2014qba} proposed an ansatz in position space that maps a linearised fluctuation in the gravity theory to a convolution of two fields, one from each gauge theory of the double copy. Thus the linearised fluctuation of a gravitational field configuration,  $\Phi_G$, will have a double copy description,
\begin{equation}
\Phi_G(x) = \left[ \phi \star\tilde{\phi} \right] (x) = \int \phi(y) \, \tilde{\phi}(x-y) dy \,,
\label{conv-int}
\end {equation}
where the $\star$ denotes a convolution, and where $\phi$ and $\tilde{\phi}$ denote field theory configurations.
Linearised gauge transformations of the field quantities on the right hand side in the above equation result in linearised local symmetry transformations on the gravity side.  Further, differential operators acting on the double copy convolution \eqref{conv-int} are allowed to hit either of the terms in the convolution. For the model discussed in this paper, the 
field theory containing $\phi$ exhibits the same number of supersymmetries (namely ${\mathcal N}=2$) 
as the gravity theory, while the field theory containing $\tilde \phi$ carries no supercharges.
Therefore, suppose that one establishes an ansatz for any given on-shell $\Phi_G$ such that local symmetries on both sides of the double copy equality are mapped to each other. Then,
the on-shell differential operator $D$ which annihilates $\Phi_G$ to form its equation of motion
$D\Phi_G=0$, annihilates the convolution \eqref{conv-int} of the two corresponding fields in the double copy via
the field theory equations of motion.
Under 
a linearised supersymmetry transformation acting on the gravitational field, the supersymmetry transformation also acts on the corresponding field in the supersymmetric theory on the double copy side. On the gravity side, under this supersymmetry transformation one gets a new on-shell field.
  The same is true for the supersymmetric field theory in the double copy. Hence, the newly produced on-shell field configurations in the gravity and gauge sectors must be  mapped to each other by a consistent dictionary. This must be true for all states that are generated down the supersymmetry ladder, and hence one expects the dictionary to be consistently established for all on-shell fields. 
In this paper we check this argument for linearised ${\mathcal N}=2$ supergravity coupled to one vector multiplet
with prepotential $F = - i X^0 X^1$.

The idea of developing a double copy dictionary for non-supersymmetric gravitational solutions has been pursued 
recently in \cite{Monteiro:2014cda,Luna:2015paa,Luna:2016due,White:2016jzc}. In this note we will address a similar question for gravitational BPS solitons.

 We start with a double copy  ansatz for a combination of the gravitini and gaugini. 
By repeated application of the linearised supersymmetry transformations, we derive the on-shell dictionary. This dictionary is valid for all long multiplet configurations.  However, for short (BPS) configurations, which have vanishing fermionic fields, the full dictionary cannot be generated by supersymmetry transformations in the procedure
described above. Hence, the proof of the double copy prescription is a priori not valid for such states. 
Also, these BPS configurations are sourced, while the dictionary we construct is for on-shell source free configurations. However, this mirrors the case of on-shell linearised  supergravity theories which are constructed in source free set-ups, but whose equations of motion generate sourced solitonic states. So, the validity of this dictionary for BPS states cannot be trivially discounted.
 Therefore, we test the dictionary  empirically on these states by applying it to the  weak-field approximation of dyonic (carrying both electric and magnetic charges) BPS black holes in this theory, and find that it holds. Hence we conjecture that the dictionary holds for
 all linearised BPS on-shell configurations in this model.
 We conclude with some comments on technical caveats in the dictionary and point out the next steps in the program. 

\section{Double copy dictionary}\label{Dictionary}

We construct the on-shell double copy dictionary for one of the simplest four-dimensional $\mathcal{N}=2$ supergravity
theories with vector multiplets, namely supergravity coupled to one vector multiplet based on the prepotential
$F = - i X^0 X^1$. We use the superconformal approach to $\mathcal{N}=2$ supergravity 
\cite{deWit:1979dzm,deWit:1980lyi,deWit:1984wbb,deWit:1984rvr,Bergshoeff:1980is}. A brief summary of some of its features can be found in \autoref{appendix_conventions}.

Traditionally, the on-shell double copy dictionaries given in the literature are in terms of momentum states (see \cite{Bianchi:2008pu,BjerrumBohr:2010rt,Damgaard:2012fb,Carrasco:2012ca,Carrasco:2013ypa,
Chiodaroli:2014xia,Anastasiou:2015vba,Anastasiou:2013hba} for some examples). Here, we will derive a double copy dictionary in position space by means of the convolution \eqref{conv-int}.

For the model at hand, the double copy construction proceeds by tensoring an $\mathcal{N}=2$ super Yang-Mills multiplet with an ($\mathcal{N}=0$) gauge field. At the level of momentum states, it was shown in 
\cite{Anastasiou:2015vba} that these are the multiplets that are relevant for the double copy construction of
this model. This is displayed in \autoref{table_on_shell_states}, where we give the helicity eigenstates that result from the tensoring. Since the Yang-Mills fields may lie in a representation of a global non-Abelian group, an additional spectator
field will have to be included in the dictionary, leading to a generalization of \eqref{conv-int} 
\cite{Anastasiou:2014qba}.

\begin{table}[h]
\small
\begin{center}
 $\begin{array}{c|c|c}
 &\begin{array}{c}  \tilde{A}^-\end{array}& \begin{array}{c} \tilde{A}^+\end{array}\\
\hline
&&\\
\begin{array}{c}{A}^-\end{array}
&\begin{array}{cccccc} g^- &\end{array} 
&\begin{array}{cccccc}  \varphi_0\end{array}
\\

&&\\
\begin{array}{c}\lambda_i^- \end{array}
&\begin{array}{cccccc} \psi_i^{-}\end{array} 
&\begin{array}{cccccc} \chi_i^+\end{array}
\\

&&\\
\begin{array}{c} \sigma^+,\sigma^- \end{array}
&\begin{array}{cccccc} A_{0,1}^-\end{array} 
&\begin{array}{cccccc} A_{0,1}^+\end{array}
\\

&&\\
\begin{array}{c}\lambda_i^+ \end{array}
&\begin{array}{cccccc} \chi_i^-\end{array} 
&\begin{array}{cccccc} \psi_i^{+}\end{array}
\\

&&\\
\begin{array}{c}{A}^+\end{array}
&\begin{array}{cccccc}   \varphi_1\end{array}
&\begin{array}{cccccc} g^+ &\end{array} 

\\

\end{array}$
\caption{On-shell $(\mathcal{N}=2)_{SYM}\times(\mathcal{N}=0)_{SYM}= (\mathcal{N}=2)_{sugra}+(\mathcal{N}=2)_{SYM}$}
\label{table_on_shell_states}
\end{center}
\end{table} 
\noindent

The double copy dictionary is a dictionary for fluctuations around a fixed background.  On the supergravity
side, we take the background to be given by flat spacetime, allowing for the presence of constant scalar fields
which we denote by $\langle X^I \rangle$ ($I = 0,1$). On the super Yang-Mills side, the background is also 
taken to be flat spacetime.
We then derive the double copy dictionary by linearising these supersymmetric theories around these
backgrounds. 
To keep the local symmetries manifest, 
we work with the corresponding gauge invariant quantities, i.e. field strenghts, on the gravity side, to exhibit
the double copy dictionary for these quantities.

In the following, we begin by reviewing the convolution structure in the presence of the aforementioned spectator
field, and we discuss restrictions on the convolution integrals imposed by the equations of motion.
Next, we display the linearised supersymmetry transformation rules that we will use to generate
the double copy dictionary for all the fields involved.  Then, we proceed to explain our double copy ansatz.
Finally, we use the linearised supersymmetry transformation laws to work out the double copy relations
for the supergravity fields. 
We verify
that the linearised supersymmetry transformations on the super Yang-Mills side reproduce the linearised supergravity
transformation rules.  We refer to \autoref{dderiva} for a detailed derivation of the double copy dictionary.
Our on-shell dictionary is summarized in \eqref{the_big_dictionary}.

\subsection{Convolution structure}

Following \cite{Anastasiou:2014qba}, we allow the two fields that appear in the convolution integral 
\eqref{conv-int} to transform in the adjoint representation of 
non-Abelian global groups $G$ and $\tilde{G}$, respectively. Since the 
supergravity fields we will obtain through the double copy construction do not transform under these global transformations, 
a bi-adjoint spectator field $\phi_{a\tilde{a}}$ will have to be introduced into \eqref{conv-int} so as obtain a combination
that is inert under global $G$ ($\tilde G$) transformations. Thus, rather than working with \eqref{conv-int},
we will base our dictionary on the convolution structure  \cite{Anastasiou:2014qba}
\be
\varphi_{sugra}=\varphi_{SYM}^a\star\phi_{a\tilde{a}}\star\tilde{\varphi}^{\tilde{a}}_{YM} \;,
\label{star2}
\ee
where the indices $a, \tilde a$ denote adjoint indices.
The real scalar $\phi_{a\tilde{a}}$ transforms in the bi-adjoint of $G\times\tilde{G}$,
\be
\delta\phi_{a\tilde{a}}=-f^b_{\ ac}\phi_{b\tilde{a}}\theta^c 
-f^{\tilde{b}}_{\ \tilde{a}\tilde{c}}\phi_{a\tilde{b}}\theta^{\tilde{c}} \;.
\label{phi-spec-transf}
\ee 
This scalar also appeared in the context of double copies in scattering amplitudes in \cite{Hodges:2011wm,Cachazo:2013iea} and in supergravity solutions in \cite{Monteiro:2014cda,Luna:2015paa,Luna:2016due}.
In addition to these global transformations, the (super) Yang-Mills gauge fields in \eqref{star2} also transform
under local Abelian gauge transformations with parameters $\alpha^a (x)$ and ${\tilde \alpha}^{\tilde a} (x)$, respectively.

In \eqref{star2}, $\star$ denotes the convolution
\be
[f\star g](x)=\int d^4y f(y)g(x-y) \;.
\ee
This is an associative operation, which doesn't satisfy the Leibniz rule, but instead satisfies
\be
\partial_\mu(f\star g)=(\partial_\mu f)\star g=f\star(\partial_\mu g) \;.
\label{shift-der}
\ee
We will make extensive use of this property\footnote{Note that \eqref{shift-der} holds in Cartesian coordinates, and hence we will present
the double copy dictionary in these coordinates.}  when imposing equations of motion on both sides of the double copy
relation \eqref{star2}, as well as when checking the transformation behaviour of both sides under linearised
supersymmetry.
Specifically,  we will find that when imposing equations of motion on \eqref{star2}, we are led to
constraints of the form
\begin{equation}
\partial^{\mu}  \left( \varphi_{SYM}^a\star\phi_{a\tilde{a}}\star\tilde{A}_{\mu}^{\tilde{a}} \right) =0 \;,
\label{star-constr}
\end{equation}
where $\varphi_{SYM}^a$ is composed of fields from the ${\mathcal N}=2$ super Yang-Mills multiplet.
Using \eqref{shift-der}, this equals 
\begin{equation}
\varphi_{SYM}^a\star\phi_{a\tilde{a}}\star \partial^{\mu} \tilde{A}_{\mu}^{\tilde{a}}  =0 \;,
\end{equation}
which is automatically satisfied if we work in the Lorentz like gauge
\begin{equation}
\partial^{\mu}  \tilde{A}_{\mu}^{\tilde{a}}  =0 \;.
\label{lorgaugea}
 \end{equation}
 Under a local Abelian transformation ${\tilde A} \rightarrow {\tilde A} + d \tilde{\alpha}$, we find
 the restriction (from \eqref{star-constr})
 \begin{equation}
\varphi_{SYM}^a\star\phi_{a\tilde{a}}\star \Box \tilde{\alpha}^{\tilde{a}}  =0 \;,
\end{equation}
which becomes $\Box \tilde{\alpha}^{\tilde{a}}  =0$ in the gauge \eqref{lorgaugea}.

For simplicity, and without loss of generality, we will develop the double copy dictionary 
in the gauge \eqref{lorgaugea}.

\subsection{Linearised transformation laws}

In a double copy relation such as \eqref{star2}, subjecting fields on one of the sides to a transformation, induces
a transformation of the fields on the other side. Thus, subjecting the fields on the right hand side
to supersymmetry transformations will induce a supersymmetry transformation of the supergravity fields on the left hand side, and vice versa.  
This will be exploited below to construct the double copy dictionary for the supergravity theory based on 
the prepotential $F = -i X^0 X^1$.

Thus, let us display the linearised transformation laws which we will be using in the following when setting
up the double copy dictionary. Consider first a $\mathcal{N}=2$ super Yang-Mills multiplet 
lying in a representation
of a global non-Abelian group $G$. It transforms as follows under rigid supersymmetry ($\epsilon$), 
local Abelian ($\alpha (x)$) and
global non-Abelian ($\theta$) transformations,
\be 
\label{rigid-transf}
\begin{aligned}
\delta A_\mu^a&=\left( \frac{1}{2}\varepsilon^{ij}\bar{\epsilon}_i\gamma_\mu\lambda_j^a+h.c. \right) +\partial_\mu\alpha^a+ f^a_{\ bc}A_\mu^b\theta^c   \;, \\
\delta\lambda_i^a&=\gamma^\mu\partial_\mu\sigma^a\epsilon_i +\frac{1}{4}\gamma^{\mu\nu}F^{a-}_{\mu\nu}\varepsilon_{ij}\epsilon^j +f^a_{\ bc}\lambda_i^b\theta^c   \;, \\
\delta\sigma^a&=\frac{1}{2}\bar{\epsilon}^i\lambda^a_i+f^a_{\ bc}\phi^b\theta^c \;,
\end{aligned}
\ee
where $f^a_{\ bc}$ denote the structure constants of the global non-Abelian group $G$.

Note that the bosonic transformations above can be seen as the linearisation of the transformations corresponding to a local non-Abelian gauge group \cite{Anastasiou:2014qba}. In this sense, we are mapping linearised super Yang-Mills theory to linearised supergravity.  

An ${\mathcal N}=0$ gauge field transforms as
\be
\delta \tilde{A}_\mu^{\tilde{a}}=\partial_\mu\tilde{\alpha}^{\tilde{a}} +\tilde{f}^{\tilde{a}}_{\ \tilde{b}\tilde{c}} \tilde{A}_\mu^{\tilde{b}}\tilde{\theta}^{\tilde{c}}  \;,
\ee
where the global non-Abelian group $\tilde{G}$ (with structure constants $\tilde{f}^{\tilde{a}}_{\ \tilde{b}\tilde{c}}$)
may be different from the global group $G$ above. 
Using these transformations laws together with \eqref{phi-spec-transf}, the convolution \eqref{star2} is indeed inert
under global $G \times {\tilde G}$ transformations.

Next, let us consider the linearised supersymmetry transformation rules for the fields appearing in the ${\cal N}=2$
supergravity theory based on the prepotential $F = - i X^0 X^1$. As mentioned above, we linearise around a flat spacetime background with metric $\eta_{\mu \nu}$ and constant scalar fields $\langle X^I \rangle $, and hence, we linearise
the spacetime metric and the scalar fields as
\begin{eqnarray}
g_{\mu \nu} &=& \eta_{\mu \nu} + h_{\mu \nu} \;, \nonumber\\
X^I &=& \langle X^I \rangle + \delta X^I \;.
\label{flatbackg}
\end{eqnarray}
The physical scalar field $z = X^1/X^0$ is then linearised as
\begin{equation}
z = \frac{X^1}{X^0} = \langle z \rangle + \delta z \;\;\;,\;\;\;  \langle z \rangle = 
 \frac{\langle X^1 \rangle}{\langle X^0 \rangle} \;\;\;,\;\;\; \delta z = \frac{1}{\langle X^0 \rangle}
 \left(\delta X^1 - \langle z \rangle \, \delta X^0 \right) \;.
\end{equation}
In the following, in order to avoid cluttering of notation, we will denote the fluctuations $\delta X^I$ simply by
$X^I$.

Linearising the supergravity transformation rules summarized in 
\autoref{App:Linearising the Supergravity Transformation Rules}, we obtain for the model based on $F = -i X^0 X^1$ the following
Q-supersymmetry transformation rules 
(dropping pure gauge terms in the variation of the gravitini),
\vspace{-0pt}  
\be
\label{all_sugra_SUSY}
\begin{aligned}
\delta_Q h_{\mu\nu}&=\bar{\epsilon}^i\gamma_{(\mu}\psi_{\nu)i}+h.c.   \;, \\
\delta_Q \psi_\mu^i&=-\frac{1}{4}\gamma^{ab}\partial_{[a}h_{b]\mu}^-\epsilon^i-\frac{1}{16}T_{\alpha\beta}^-\gamma^{\alpha\beta}\gamma_\mu\varepsilon^{ij}\epsilon_j  \;, \\
\delta_Q W_\mu^0&=\frac{1}{2}\varepsilon^{ij}\bar{\epsilon}_i\gamma_\mu\Omega^0_j  
+\varepsilon^{ij}\bar{\epsilon}_i\psi_{\mu j}\langle X^0\rangle +h.c. \;, \\
\delta_Q W_\mu^1&=-\frac{\langle\bar{z}\rangle}{2}\varepsilon^{ij}\bar{\epsilon}_i\gamma_\mu\Omega^0_j 
+\langle z\rangle\varepsilon^{ij}\bar{\epsilon}_i\psi_{\mu j}\langle X^0\rangle +h.c.    \;, \\
\delta_Q\Omega^{0i}&=\gamma^\mu\partial_\mu\bar{X}^0\epsilon^i +\frac{1}{4}\gamma^{\mu\nu}
\mathcal{F}_{\mu\nu}^{0+}\varepsilon^{ij}\epsilon_j    \;, \\
\delta_Q X^I&=\frac{1}{2}\bar{\epsilon}^i\Omega_i^I \;,
\end{aligned} 
\ee  
where we used the gauge fixing condition for S-supersymmetry,
\begin{equation}
\Omega_i^1 = - \langle \bar z \rangle \, \Omega^0_i \;,
\label{SOm10}
\end{equation}
to express $\Omega_i^1$ in terms of $\Omega^0_i$. In the above, $\pm$ denote the (anti)selfdual parts, and the composite quantities 
$T_{\mu\nu}^-$ and $\mathcal{F}_{\mu\nu}^{0+}$ are given by
\be
\label{combs_field_strengths}
\begin{aligned}
T_{\mu\nu}^-&=\frac{1}{\langle\bar{X}^0\rangle}\left[ F_{\mu\nu}^{0-}+\frac{F_{\mu\nu}^{1-}}{\langle\bar{z}\rangle}\right]   \;, \\
\mathcal{F}_{\mu\nu}^{0+}&=\frac{1}{2}\left[F_{\mu\nu}^{0+}-\frac{F_{\mu\nu}^{1+}}{\langle z\rangle} \right]  \;.
\end{aligned} 
\ee

\subsection{Dictionary}

Now we derive the on-shell double copy dictionary for linearised supergravity based on $F=-i X^0 X^1$.
The dictionary is summarized in \eqref{the_big_dictionary}.

To avoid ambiguities arising from gauge degrees of freedom when going on-shell, we 
will work with field strengths in our dictionary.  Thus, rather than working with the metric fluctuation $h_{\mu \nu}$ we will work with the 
linearised Riemann tensor,
\begin{eqnarray}
R_{\rho \sigma \mu \nu} = - 2 \, \partial_{[\mu} \partial_{[\rho} h_{\sigma] \nu]} 
\:,
\label{Rlin}
\end{eqnarray}
which  is invariant under linearised diffeomorphisms,
\begin{equation}
\delta h_{\mu \nu} = \partial_{\mu} \xi_{\nu} + \partial_{\nu} \xi_{\mu} \;.
\end{equation}
Similarly, we work with $\psi^i_{\mu\nu}=2\partial_{[\mu}\psi^i_{\nu]}$ for the gravitini, and so on.

Our strategy consists of postulating the following double copy ansatz for a linear combination of the supergravity fermions, 
\be 
a\psi_{\mu\nu}^i+2b\gamma_{[\nu}\partial_{\mu]}\Omega^{0i}
\equiv \varepsilon^{ij}\lambda_j^a\star\phi_{a\tilde{a}}
\star\tilde{F}_{\mu\nu}^{\tilde{a}} \;,
\label{fermdicti}
\ee
where $a, b \in \mathbb{C}$ denote complex constants.  Note that in view of \eqref{SOm10}, 
 the left hand side of \eqref{fermdicti} captures all the relevant fermionic supergravity degrees of freedom. Inspection of  \autoref{table_on_shell_states} shows that this ansatz is the most general one compatible with \autoref{table_on_shell_states}.

At this point we recall that the fermionic fields appearing in \eqref{fermdicti}
  carry different weights under the $U(1)$ subgroup of the R-symmetry group (the so-called chiral weights). The left-handed gravitino $\psi_\mu^i$ has a chiral weight that differs by one unit from the chiral weight 
of the right-handed gaugino $\Omega^{0i}$. This implies that we assign a zero chiral weight to $a$, while $b$ will have to carry a compensating chiral weight so as to make $b \, \Omega^{0i}$ have the same chiral weight as $a\psi_{\mu}^i$
and $\lambda_j^a$.

Since, in the following, the spectator field $\phi_{a\tilde{a}}$ will only play a passive role, we will 
omit its presence,  to keep the expressions as simple as possible, and only reinstate its dependence at the end.
Thus, we will for the time being suppress the non-Abelian indices and work with the double copy ansatz 
\be 
a\psi_{\mu\nu}^i+2b\gamma_{[\nu}\partial_{\mu]}\Omega^{0i}
\equiv \varepsilon^{ij}\lambda_j 
\star\tilde{F}_{\mu\nu} \;.
\label{fermdictisimp}
\ee

By contracting \eqref{fermdictisimp} with $\gamma^{\mu}$ and using the equations of motion for the $\lambda_i$ and $\psi^i_{\mu}$ (see \autoref{dderiva}) as well as the property \eqref{shift-der},
we extract the dictionary for $\Omega^{0i}$, 
\begin{equation}
2 b \, \partial_{\nu} \Omega^{0i} = \varepsilon^{ij} \, \gamma^{\mu} \lambda_j \star \partial_{\nu} {\tilde A}_{\mu} \;,
\label{dcrOm0}
\end{equation}
and using this in \eqref{fermdictisimp} we infer the dictionary for the gravitini field strength, 
\be
2a \, \partial_{[\mu}\psi_{\nu]}^i=\varepsilon^{ij} \gamma^\rho\gamma_{[\nu}\lambda_j\star\partial_{\mu]}\tilde{A}_\rho\;.
\label{gravi-der-dcr}
\ee
These dictionary expressions have to be consistent with the linearised equations of motion for $ \Omega^{0i} $ and for the gravitini. 
This is indeed the case, as we show in \autoref{dderiva}.

Next, we return to \eqref{fermdictisimp} and verify its consistency with the equation of motion for  ${\tilde A}_{\mu}$,  by acting with $\partial^{\mu}$ on \eqref{fermdictisimp}.
This results in 
\begin{equation}
a \left( \Box \psi^i_{\nu} - \partial_{\nu} \partial^{\mu} \psi^i_{\mu} \right) = 0 \;,
\label{condeomgr}
\end{equation}
where we used the equation of motion for $\Omega^{0i}$. To verify that \eqref{condeomgr} vanishes, we take the equations
of motion 
 for the gravitini in the form $\gamma^{\mu} \, \partial_{[\mu} \psi_{\nu]}^i =0$,
 and contract it with $\gamma^\rho\partial_\rho$,
\be
\begin{aligned}
0&=\gamma^\rho\gamma^\mu[\partial_\rho\partial_\mu\psi_\nu^i-\partial_\rho\partial_\nu\psi_\mu^i]\\
&=\gamma^{\rho\mu}[\partial_\rho\partial_\mu\psi_\nu^i-\partial_\rho\partial_\nu\psi_\mu^i]
+\square\psi_\nu^i-\partial^\mu\partial_\nu\psi_\mu^i\\
&=\square\psi_\nu^i-\partial_\nu\partial^\mu\psi_\mu^i \;,
\end{aligned}
\ee
where to get to the last line we used a consequence of the gravitini equation of motion, namely $\gamma^{\nu \rho} \partial_{\nu} \psi_{\rho}^i = 0$.

Next, we apply supersymmetry transformations to the double copy relations \eqref{dcrOm0} and \eqref{gravi-der-dcr}, to infer the double copy relations
for the remaining supergravity fields. This will be discussed at length in \autoref{dderiva}, to which we refer.
For the combinations given in \eqref{combs_field_strengths} we obtain the double copy relations
\begin{eqnarray}
a \, T^-_{\mu \nu} &=& - 4  \sigma \star {\tilde F}^-_{\mu \nu} \;, \nonumber\\
b \, \mathcal{F}_{\mu\nu}^{0+} 
&=& -  \sigma \star {\tilde F}^+_{\mu \nu} \;, 
\end{eqnarray}
and hence
\begin{eqnarray}
F^{0 +} &=& - \frac{1}{b}\, \sigma \star {\tilde F}^+ - \frac{2 \langle X^0 \rangle }{\bar a} \, {\bar \sigma} \star {\tilde F}^+ \;, \nonumber\\
F^{1 +} &=&  \frac{\langle z \rangle }{b}\, \sigma \star {\tilde F}^+ - \frac{2 \langle X^1  \rangle }{\bar a} \, {\bar \sigma} \star {\tilde F}^+ \;.
\label{dcrF01}
\end{eqnarray}
This yields the following relations for the supergravity field strengths $F_{\mu \nu}^I$, 
\begin{eqnarray}
F^0_{\mu \nu} &=& 
-   \left( \frac{\langle {\bar X}^0 \rangle}{a} + \frac{1}{2b} \right)
\sigma \star {\tilde F}_{\mu \nu} -  \left(\frac{1}{2b} - \frac{ \langle {\bar X}^0 \rangle}{a} \right)  \sigma \star
(^* {\tilde F} )_{\mu \nu} 
 + h.c. \;, \nonumber\\
 F^1_{\mu \nu} &=& 
  \left( - \frac{\langle {\bar X}^1\rangle }{a} + \frac{\langle z \rangle}{2b} \right)
\sigma \star {\tilde F}_{\mu \nu} + \left(\frac{\langle z \rangle }{2b} +\frac{\langle  {\bar X}^1 \rangle}{a} \right)  \sigma \star
(^* {\tilde F})_{\mu \nu}
 + h.c. \;.
 \label{dcF01}
\end{eqnarray}
Note that the expressions for $F^0$ and $F^1$ get interchanged under
\begin{equation}
\langle X^0 \rangle \leftrightarrow \langle X^1 \rangle \;\;\;,\;\;\; \frac{1}{b} \leftrightarrow - \frac{\langle z \rangle }{b} \;.
\label{b01}
\end{equation}
More generally, we note that the symplectic transformation $\langle X^0 \rangle  \rightarrow \kappa \, \langle X^1 \rangle, \, \langle X^1\rangle \rightarrow 
\langle X^0\rangle / \kappa$, together with $1/b \leftrightarrow - \langle z \rangle \, \kappa/ b$ (with $\kappa \in \mathbb{R}$), interchanges  $F^0$ and $F^1$
and preserves the prepotential $F = -i X^0 X^1$.

We now summarize the resulting on-shell double copy dictionary for all the supergravity fields. Reinstating the dependence on the spectator field $\phi_{a\tilde{a}}$, it is given by
\be
\label{the_big_dictionary}
\boxed{
\begin{aligned}
a \, R_{\mu\nu\alpha\beta}^-&=
-\frac{1}{2}\left[F_{\mu\nu}^a\star\phi_{a\tilde{a}} \star \tilde{F}_{\alpha\beta}^{\tilde{a}-} +F_{\alpha\beta}^{a-}\star\phi_{a\tilde{a}}\star \tilde{F}_{\mu\nu}^{\tilde{a}}
-4\eta_{[\alpha[\mu}\partial_{\nu]}\partial_{\beta]}^-A^{a\rho}\star\phi_{a\tilde{a}}\star\tilde{A}_\rho^{\tilde{a}}\right]  \\
a \, \psi_{\mu\nu}^i&=\varepsilon^{ij} \gamma^\rho\gamma_{[\nu}\lambda_j^a\star\phi_{a\tilde{a}} \star\partial_{\mu]}\tilde{A}_\rho^{\tilde{a}} \\
a \, T_{\mu\nu}^-&=-4\sigma^a\star\phi_{a\tilde{a}}\star\tilde{F}_{\mu\nu}^{\tilde{a}-} \\
b \, \mathcal{F}_{\mu\nu}^{0+}&=-\sigma^a\star\phi_{a\tilde{a}} \star\tilde{F}_{\mu\nu}^{\tilde{a}+}  \\
b \, \partial_\mu\Omega^{0i}&=\frac{1}{2}\varepsilon^{ij}\gamma^\rho\lambda_j^a\star\phi_{a\tilde{a}} \star\partial_\mu\tilde{A}_\rho^{\tilde{a}}  \\
b \, \partial_\mu\bar{X}^0&=\frac{1}{2} F_{\mu\rho}^{a -}  \star\phi_{a\tilde{a}} \star\tilde{A}^{\tilde{a} \rho}
\end{aligned}
} 
\ee 
We note that in the expression for the Riemann tensor, the anti self-dual part is taken over the indices $\alpha\beta$. A completely equivalent expression is, of course, obtained if, instead, we take the anti self-dual part over $\mu\nu$.
The double copy relation for $\Omega^{1i}$ follows from the one for $\Omega^{0i}$ by virtue of the relation \eqref{SOm10}, and the double copy relations for $F_{\mu \nu}^{\pm I}$ are as in 
\eqref{dcF01}, with the spectator field reinserted. Similarly, the dictionary for $\partial_\mu X^1$ follows immediately from that for $\partial_\mu X^0$, when we use \eqref{aconnvan}. Observe that the on-shell dictionary \eqref{the_big_dictionary} is invariant under local Abelian transformations $A \rightarrow A + d \alpha , \, {\tilde A}  \rightarrow {\tilde A} + d \tilde {\alpha}$
by virtue of the equations of motion $\partial^{\mu} F_{\mu \nu} = 0$ and $\Box {\tilde \alpha} = 0$, which
follows from the Lorentz gauge condition.

We note that the expression for the Riemann tensor can also be written as
\be
aR_{\mu\nu\alpha\beta}^-=
2\left[F^a_{[\alpha [\mu}\star\phi_{a\tilde{a}} \star \tilde{F}^{\tilde{a}}_{\nu ] \beta]} 
+\eta_{[\alpha[\mu}\partial_{\nu]}\partial_{\beta]}A^{a\rho}\star\phi_{a\tilde{a}} \star\tilde{A}^{\tilde{a}}_\rho\right]^- \;.
\ee
If we restrict the parameter $a$ to $a=a_\R\in\R$, then we are left with a simpler expression for the Riemann tensor,
\be
\label{arealR}
a_\R \, R_{\mu\nu\alpha\beta}=
-\frac{1}{2}\left[F_{\mu\nu}^a\star\phi_{a\tilde{a}} \star \tilde{F}_{\alpha\beta}^{\tilde{a}} +F_{\alpha\beta}^{a}\star\phi_{a\tilde{a}}\star \tilde{F}_{\mu\nu}^{\tilde{a}}
-4\eta_{[\alpha[\mu}\partial_{\nu]}\partial_{\beta]}A^{a\rho}\star\phi_{a\tilde{a}}\star\tilde{A}_\rho^{\tilde{a}}\right] \;.
\ee

It can be checked that the supersymmetry variation of the right hand side of \eqref{the_big_dictionary}
 correctly induces the supergravity transformation of the left hand side, and vice-versa, by means of the double copy dictionary.
The double copy relations in \eqref{the_big_dictionary} are also consistent with the equations of motion of all the fields involved. Hence,
we have established a consistent on-shell double copy dictionary for this model.

The on-shell double copy dictionary is formulated in terms of field strengths. However, one could give a double copy
prescription in terms of fields, but in doing so one must be very careful in making consistent gauge choices,
in particular, when peeling off derivatives in \eqref{the_big_dictionary}.

\section{Dyonic BPS black hole solutions}

In the following we consider dyonic
BPS black hole solutions in the model $F= - i X^0 X^1$, and we verify that they have a double copy description based
on the double copy relations given in \eqref{the_big_dictionary}.

On the supergravity side, the BPS conditions are derived by imposing the restriction \cite{Behrndt:1997ny}
\begin{equation}
k \, \epsilon_i = \varepsilon_{ij} \, \gamma_0 \, \epsilon^j \;,
\label{BPSstatic}
\end{equation}
where $k$ denotes a phase factor with an appropriate chiral weight, so that both sides have the same chiral weight.
We impose the same condition on the field theory side. 

To keep the expressions as simple as possible, we again omit non-Abelian indices $(a, \tilde a)$, 
and only reinstate their dependence at the end.

\subsection{Field theory side}

On the field theory side, we seek static BPS solutions to the rigid supersymmetry transformations displayed in \eqref{rigid-transf},
\be
\begin{aligned}
\delta_Q\lambda_i&=\gamma^\mu\partial_\mu\sigma \epsilon_i +\frac{1}{4}\gamma^{\mu\nu}F^-_{\mu\nu}\varepsilon_{ij}\epsilon^j=0 \;,
\end{aligned} 
\ee
where we have suppressed the non-Abelian index $a$, as mentioned above. We impose the BPS condition \eqref{BPSstatic}, which results
in 
\be 
\begin{aligned}
\partial_0 (\sigma \bar{k})&=0 \;, \\
F_{mn}^{-}+4\partial_{[m}^-(\sigma \bar{k})\eta_{n]0}&=0 \;\;\;,\;\;\; m, n = 1,2,3 \;, 
\end{aligned}
\ee
where the superscript '$-$' denotes the anti-selfdual part. The second equation results in
\be
F_{mn}+2\left[2\partial_{[m}Re(\sigma \bar{k})\eta_{n]0}-\varepsilon_{mnp0}\partial^p Im(\sigma \bar{k}) \right]=0 \;.
\ee
We restrict to static BPS configurations supported by electric fields only. In Cartesian coordinates, the BPS configurations we consider are thus described by 
\be
\label{BPS_solss_SYM}
\begin{aligned}
\partial_t(\sigma \bar{k}) & =0 \;, \\
\partial_i Im(\sigma \bar{k})&=0 \;,\\
F_{ti} &= - 2 \partial_i Re(\sigma \bar{k})  \;. \\
F_{ij} &=  0 \;\;\;,\;\;\; i, j= x, y, z \;.
\end{aligned} 
\ee
In a spherically symmetric context, the electric potential $Re(\sigma \bar{k})$ will only depend on the radial coordinate $r$.
When comparing with the supergravity BPS solutions, we will find it convenient to work with coordinates $(u, r, \theta, \phi)$, with $u$ given by $u = t + r$.
The resulting metric $\eta_{\mu \nu}$ and its inverse $\eta^{\mu \nu}$ are given in \eqref{flat_metric_finkle}.  The gauge potential $A_u$ reads
\begin{equation}
A_u = 2 Re(\sigma \bar{k}) \;.
\label{Ausig}
\end{equation}
Remarkably, this electric BPS configuration will be mapped to a dyonic BPS black hole configuration in the following.

\subsection{Supergravity side}

The model $F= -i X^0 X^1$ admits dyonic BPS black hole solutions, as first pointed out in \cite{Ferrara:1996dd}. These solutions, briefly reviewed in \autoref{App:Linearising Supergravity Solutions}, are supported by two electric
charges, $q_0$ and $q_1$, and by two magnetic charges $p^0$ and $p^1$.  We find it convenient to work in Eddington-Finkelstein type coordinates $(u, r, \theta, \phi)$, defined in  \eqref{finkel}. We linearise the solution around a flat background of the form \eqref{flatbackg}, where $\eta_{\mu \nu}$ is given by \eqref{flat_metric_finkle}, and the background
scalar fields are 
\begin{equation}
\langle X^0 \, \bar k \rangle =  - \frac12 \, \left( h_1 - i h^0 \right)
 \;\;\;,\;\; \langle X^1 \, \bar k \rangle =  - \frac12 \, \left( h_0 - i h^1 \right) \;,
 \;\;
 \langle z \rangle =  \frac{h_0-i h^1 }{h_1- i h^0}  \;,
\end{equation}
where $h_0,h_1, h^0, h^1$ are constants entering the definition of the harmonic functions appearing in the attractor equations \eqref{attractor_equations}. 
As shown in \autoref{App:Linearising Supergravity Solutions}, the fluctuating fields are then given by
\be
\label{sugra_solution} 
\begin{aligned}
h_{\mu\nu}&=diag \, (\frac{Q}{r},0,Q r,Q r sin^2\theta) \;,\\
F^0_{ur} &=   \frac{Q h_1 - q_1}{ r^2} \;\;\;,\;\;\;
F^1_{ur} =  \frac{Q h_0 - q_0}{r^2} \;\;\;,\;\;\; F_{\theta \phi}^I = p^I \, \sin \theta \;\;\;,\;\;\; I=0,1 \;, \\
X^0 \, {\bar k} &= - \frac{  \left(q_1  -i p^0 - \tfrac12 Q ( h_1 - i h^0) \right)}{2 r}  \;,
\end{aligned}
\ee
where
\be
Q=h_0q_1+h_1q_0+ h^0 p^1 + h^1 p^0 \;.
\ee
The expression for the fluctuation $X^1 \, {\bar k}$ follows from the one for $X^0 \, {\bar k}$ by interchanging the indices $0$ and $1$, and is related to $X^0 \, {\bar k}$ by \eqref{aconnvan}.

\subsection{Double copy expressions}

Now we show that the supergravity fluctuations \eqref{sugra_solution} have a double copy description
based on the dictionary \eqref{the_big_dictionary} by determining the associated gauge field configuration
satisfying \eqref{BPS_solss_SYM}.
To this end, we will employ \eqref{arealR} (we thus take the parameter $a$ to be real) to 
write down a gauge fixed double copy expression for 
the fluctuating metric $h_{\mu \nu}$
(suppressing non-Abelian indices)
\begin{equation}
a \, h_{\mu \nu} = A_{\mu} \star {\tilde A}_{\nu} + A_{\nu} \star {\tilde A}_{\mu} -  \left(A_{\alpha} \star {\tilde A}^{\alpha}
\right) \eta_{\mu \nu} \;.
\label{hdcprel}
\end{equation}
Note that the double copy relations were derived in the Cartesian coordinate system.
Here, we work with spherical coordinates $(u, r, \theta, \phi)$, for convenience. In this coordinate system, the convolution integral for fields reads
\begin{equation}
\left[ \phi \star\tilde{\phi} \right] (x) = \int \phi(y) \, \tilde{\phi}(x-y) \sqrt{-\eta (y)} \, d^4y \,,
\end {equation}
while for field strengths we use
\be
\sigma \star {\tilde F}_{MN} = 2 \, \partial_{[M} \left( \sigma \star {\tilde A}_{N]} \right) \;.
\ee
Further, note that the convolution of a function with the dual of a tensor, $\sigma \star (^*\tilde{F})$ is implemented as the dual of the convolution of the function with the tensor, $^* ( \sigma \star \tilde{F})$, which is consistent with the Cartesian implementation.

We take the non-vanishing components of $A_{\mu}$ and ${\tilde A}_{\mu}$ to be 
\begin{eqnarray}
\label{choiceq0}
A_{u} &=& \frac{d_1}{r}  \;, \\
{\tilde A}_{u}  &=& \delta^{(4)} (u, r, \theta, \phi) \;\;\;,\;\;\; {\tilde A}_{r}  =   d_2 \, \delta^{(4)} (u, r, \theta, \phi) \;, \nonumber\\
 \delta^{(4)} (u-u_0, r-r_0 , \theta-\theta_0, \phi-\phi_0) &=& \delta (u-u_0) \delta(r-r_0) \delta (\theta-\theta_0) \delta(\phi-\phi_0)/(r^2 \sin \theta) \;,
 \nonumber
\end{eqnarray}
with constant coefficients $d_1$ and $d_2$ which we now determine. Inserting this into \eqref{hdcprel}
we find the relations
\begin{eqnarray}
a \, Q &=& d_1 \left(2 + d_2 \right) \;, \nonumber\\
a  \, Q &=& - d_1 \, d_2 \;, 
\end{eqnarray}
from which we infer
\begin{eqnarray}
d_1 &=& a \, Q \;, \nonumber\\
d_2 &=& - 1 \;.
\end{eqnarray}
Then, using \eqref{BPS_solss_SYM} and \eqref{Ausig}, we take
\begin{equation}
2 Re(\sigma \bar{k}) = \frac{d_1}{r} \;\;\;,\;\;\; Im (\sigma \bar{k}) =0 \;.
\label{relphid}
\end{equation}
This, together with \eqref{choiceq0}, determines the field configuration on the gauge theory side.
We now show that it correctly reproduces the fluctuating fields \eqref{sugra_solution}.

We first
consider the double copy description of the fluctuating scalar field $X^0$.
Using \eqref{flucX0z} we obtain
\begin{equation}
\bar k  \, \partial_{\mu}  X^0   =  - \frac{ \left(q_1  -i p^0 - \tfrac12 Q ( h_1 - i h^0) \right) }{2 }    \, \partial_{\mu} \frac{1}{r}\;.
\label{expdz}
\end{equation}
On the other hand, using the double copy relations \eqref{the_big_dictionary} for the electric field configuration
\eqref{choiceq0}, 
we infer,
\begin{equation}
\bar k \, \partial_{\nu} X^0 = \frac{\bar k }{4 \bar b} \, \partial_{\nu} \left( {A}^{\alpha} \star {\tilde A}_{\alpha}
\right) = \frac{\bar k }{4 \bar b} \, \partial_{\nu} \frac{d_1 d_2}{r} = - 
\frac{\bar k \, a \, Q}{4 \, \bar b} \, \partial_{\nu} \frac{1}{r} 
\;.
\end{equation}
Comparing with \eqref{expdz} we obtain
\begin{equation}
\frac{\bar k \, a }{ 2 \, \bar b} =  \frac{ q_1  -i p^0 - \tfrac12 Q ( h_1 - i h^0) }{ Q} \;.
\label{relbh1}
\end{equation}
Note that under the interchange of the 0-sector and the 1-sector, this transforms as in \eqref{b01}
by virtue of the BPS relation $h^I q_I = h_I p^I$, as it should.

Next we employ \eqref{relbh1} as well as \eqref{relphid} in the double copy relations for the 
field strengths $F^I_{\mu \nu}$ using \eqref{dcF01}. First we consider $F^0_{\mu \nu}$.
Using the result that the only non-vanishing integral in \eqref{dcF01} is $\sigma \star {\tilde F}_{ur}$,
we obtain
\begin{eqnarray}
F^0_{ur} &=& 
-   \left( \frac{\langle {\bar X}^0 \rangle }{a} + \frac{1}{2b} \right)
\sigma \star {\tilde F}_{ur}  + h.c.  \nonumber\\
&=& -   \left( \frac{\langle {\bar X}^0 \, k \rangle }{a} + \frac{k}{2b} \right) \partial_r
\left( \sigma \, \bar k \star \left( -  {\tilde A}_u \right) \right) + h.c.  \nonumber\\
&=& \frac{(Q h_1  - q_1)}{aQ}  \frac{d_1}{r^2} = \frac{Q h_1  - q_1}{r^2}  \;,
\nonumber\\
F^0_{\theta \phi} &=& i \varepsilon_{\theta \phi}{}^{ur} 
\left( - \frac{\langle {\bar X}^0 \rangle }{a} + \frac{1}{2b} \right)
\sigma \star {\tilde F}_{ur}  + h.c.  \nonumber\\
&=& -2 \varepsilon_{\theta \phi}{}^{ur} 
Im \Big[ \left( - \frac{\langle {\bar X}^0 k \rangle }{a} + \frac{k}{2b} \right) \Big]
\sigma \, {\bar k} \star {\tilde F}_{ur}   \nonumber\\
&=& \frac{2 \, p^0}{a \, Q} r^2 \, \sin \theta \, \partial_r
\left( \sigma \, \bar k \star \left( - {\tilde A}_u \right)  \right)  \nonumber\\
&=& \frac{p^0}{a \, Q} r^2 \, \sin \theta \, 
 \frac{d_1}{r^2} = p^0 \, \sin \theta \;,
  \end{eqnarray}
in agreement with \eqref{sugra_solution}. 

Now we consider $F^1_{\mu \nu}$ and compute
\begin{eqnarray}
Re   \left( - \frac{\langle {\bar X}^1 \, k \rangle }{a} + \frac{\langle z \, k  \rangle}{2b} \right) &=& \frac{Q \, h_0 - q_0}{a \, Q}
\;, \nonumber\\
Im \left(\frac{\langle z \, k \rangle }{2b} +\frac{\langle  {\bar X}^1 \, k  \rangle}{a} \right)  &=& - \frac{p^1}{a \, Q}\;,
 \end{eqnarray}
 \vspace{-0pt}
 where we made use of the normalization condition \eqref{normh01} as well as of the BPS constraint $h^I q_I = h_I p^I$.
 We obtain
 \begin{eqnarray} 
 F^1_{ur} &=& 2 Re 
  \left( - \frac{{\langle \bar X}^1 \, k \rangle }{a} + \frac{\langle z \, k \rangle}{2b} \right)
\sigma \, {\bar k} \star {\tilde F}_{ur}   \nonumber\\
&=&2 Re  \left( - \frac{\langle {\bar X}^0 \, z \, k \rangle }{a} + \frac{\langle z \rangle \, k}{2b} \right) \partial_r
\left(\sigma\, \bar k \star \left( -  {\tilde A}_u \right) \right) \nonumber\\
&=&  \frac{(Q \, h_0 - q_0)}{a \, Q}  \frac{d_1}{r^2} =  \frac{Q \, h_0 - q_0}{r^2} \;, \nonumber\\
F^1_{\theta \phi} &=& -  i \varepsilon_{\theta \phi}{}^{ur} 
\left( \frac{\langle {\bar X}^1 \rangle }{a} + \frac{\langle z \rangle }{2b} \right)
\sigma \star {\tilde F}_{ur}  + h.c.  \nonumber\\
&=& 2 \varepsilon_{\theta \phi}{}^{ur} 
Im \Big[ \left(  \frac{\langle {\bar X}^1 k \rangle }{a} + \frac{\langle z \rangle k}{2b} \right) \Big]
\sigma \, {\bar k} \star {\tilde F}_{ur}   \nonumber\\
&=& \frac{2 \, p^1}{a \, Q} r^2 \, \sin \theta \, 
\partial_r \left( \sigma \, \bar k \star \left( -  {\tilde A}_u \right)  \right)  \nonumber\\
&=& \frac{p^1}{a \, Q} r^2 \, \sin \theta \, 
 \frac{d_1}{r^2} = p^1 \, \sin \theta \;,
 \end{eqnarray}
in agreement with \eqref{sugra_solution}. 
Thus, we have verified that the double copy gauge field configuration
\eqref{choiceq0} and \eqref{relphid} correctly reproduces the gravitational configuration \eqref{sugra_solution}.

We may easily reinstate the dependence on the non-Abelian indices, by taking $A_{\mu}^a = A_{\mu} \, c^a, \; {\tilde A}_{\mu}^a = {\tilde A} _{\mu} \, {\tilde c}^{\tilde a}, \;
\phi_{a \tilde a} = V_{ a \tilde a} \,  \delta^{(4)} (u, r, \theta, \phi) $, with constant $c^a, {\tilde c}^{\tilde a}, V_{ a \tilde a}$ normalised to $c^a V_{ a \tilde a} {\tilde c}^{\tilde a} = 1$.

Finally, we observe that the configuration \eqref{choiceq0} satisfies the constraint \eqref{star-constr}. Hence,
we conclude that in the weak field approximation, the dyonic  BPS black hole solution has a double copy description, based on  \eqref{the_big_dictionary}, in terms of an electrically charged  BPS solution in gauge theory.

\section{Conclusions}

We have constructed the on-shell double copy dictionary for a $D=4, {\mathcal N}=2$ supergravity theory with one vector multiplet based on the prepotential $F = -i X^0 X^1$. In doing so,
we made use of the explicit S-supersymmetry gauge fixing condition \eqref{SOm10}. For a different one vector multiplet prepotential, such as $F = - (X^1)^3/X^0$,  this condition
will look different. Thus, the double copy dictionary obtained here only applies to this particular prepotential.

Note that the double copy dictionary \eqref{the_big_dictionary} was derived for source free theories. However,
we showed that it also holds for a class of gravitational BPS configurations that have sources.

An important feature of the double copy construction is that the field configurations in the $\mathcal{N}=2$ and 
$\mathcal{N}=0$ sectors
are constrained by 
\begin{equation}
\partial^{\mu}  \left( \varphi_{SYM}^a\star\phi_{a\tilde{a}}\star\tilde{A}_{\mu}^{\tilde{a}} \right) =0 \;.
\end{equation}
This relation also imposes a constraint
on the gauge transformations of the field ${\tilde A}_{\mu}^a$ in the $\mathcal{N} =0$ sector.
Although this relation appears to couple fields in the two sectors,
field configurations where $\tilde{A}_{\mu}^a$
obeys a Lorentz like gauge condition \eqref{lorgaugea}
trivially satisfy this constraint, keeping the fields in the two
sectors independent.

In this note we have also given a double copy description of all BPS single center black holes 
in the model based on the prepotential $F = -i X^0 X^1$. Most interestingly, this description maps
dyonic black holes to purely electric configurations on the field theory side. An obvious generalization consists
in introducing a dyonic configuration in the field theory side and examining the corresponding configuration in
gravity.

The on-shell double copy construction for ${\mathcal N}=2$ supergravity theories with more than one vector multiplet will bring in new ingredients.
Namely, adding vector multiplets on the supergravity side will require adding scalar fields in the 
${\mathcal N}=0$ sector of the double copy \cite{Anastasiou:2015vba}.
We plan to address this in the near future.


\subsection*{Acknowledgements}

\noindent
We would like to thank Sergei Alexandrov, Alexandros Anastasiou, Paolo Benincasa, Marco Chiodaroli, Michael Duff, Roberto Emparan, Ricardo Monteiro, Michele Zoccali  for helpful discussions.  
This work was supported by FCT/Portugal through
grant EXCL/MAT-GEO/0222/2012 (G.L.Cardoso), through a CAMGSD post-doc fellowship (S. Nagy)  and through FCT fellowship SFRH/BPD/101955/2014 (S. Nampuri).
This work was also supported by the COST action MP1210
"The String Theory Universe". G.L.C. would like to thank the Max-Planck-Institut f\"ur
Gravitationsphysik (Albert-Einstein-Institute) and the University of the Witwatersrand for kind hospitality during 
various stages of this
work.
\appendix

\section{Conventions}\label{appendix_conventions}

We follow the conventions of \cite{Freedman:2012zz}.
We usually denote spacetime indices by $\mu, \nu, \dots$, frame indices  by $a, b, \dots = 0,1,2,3$ and  $SU(2)$ R-symmetry indices by $i,j, \dots=1,2$.
We use the following (anti-)symmetrization conventions,
\begin{eqnarray}
[a,b] = \tfrac12 (ab - ba) \;\;\;,\;\;\; (ab) = \frac12 (ab + ba) \;.
\end{eqnarray}
We take
\begin{eqnarray}
\gamma_a \gamma_b = \eta_{ab} + \gamma_{ab} \;\;\;,\;\;\; \gamma_{ab} = \tfrac12 [\gamma_a, \gamma_b] \;,
\label{gam-decom}
\end{eqnarray}
where $\eta_{ab}= {\rm diag} (-, +, +, +)$. Introducing $\gamma_5 = i \gamma_0 \gamma_1 \gamma_2 \gamma_3$, we define
projection operators  in the usual way,
\be
P_L=\frac{1}{2}\left(\mathbb{I}+\gamma_5\right),\quad
P_R=\frac{1}{2}\left(\mathbb{I}-\gamma_5\right)  \;.
\ee
The chirality assignment of a chiral fermion is specified by the position of the $SU(2)$ R-symmetry index, for instance
\begin{eqnarray}
\psi_{\mu i} &=&P_R \, \psi_{\mu i} \;\;\;,\;\;\; \psi_\mu^i=P_L \, \psi_\mu^i \;, \nonumber\\
\Omega^{I i} &=& P_R \, \Omega^{I i} \;\;\;,\;\;\; \quad \Omega^I_i=P_L \, \Omega_i^I \;, \nonumber\\
\epsilon_i  &=& P_R \, \epsilon_i  \;\;\;,\;\;\; \epsilon^i  = P_L \, \epsilon^i  \;.
\end{eqnarray}
Under h.c., an $SU(2)$ R-symmetry index changes position.

The complete antisymmetric tensor $\varepsilon_{abcd}$ satisfies
$\varepsilon_{0123} = 1$.
The dual of an antisymmetric tensor field $F_{ab}$ is given by
\begin{equation}
(^*F)_{ab} = - \frac{i}{2} \,  \varepsilon_{abcd} F^{cd} \;.
\end{equation}
The (anti-)selfdual part of $F_{ab}$ is determined by
\begin{equation}
\label{appA_dual_tensors}
F^{\pm}_{ab} = \tfrac12 \left( F_{ab} \pm (^* F)_{ab} \right) \;.
\end{equation}
We note the relations
\begin{eqnarray}
\label{appA_gammas}
\gamma_{ab} &=& \frac{i}{2}  \,  \varepsilon_{abcd} \, \gamma^{cd} \, \gamma_5 \;, \nonumber\\
\gamma^{ab} F_{ab} \,  \epsilon^i &=& \gamma^{ab} F_{ab}^- \,  \epsilon^i \;,
\end{eqnarray}
as well as
\be
\label{app_anti_self_contract}
\gamma_\rho\gamma^{\alpha\beta}F_{\alpha\beta}^-(P_L\chi)=-4\gamma^\alpha F_{\alpha\rho}^-(P_L\chi) \;.
\ee

\section{Supergravity transformation rules }\label{App:Linearising the Supergravity Transformation Rules}

We work within the superconformal approach for $\N=2$ supergravity coupled to $\N=2$ vector multiplets 
\cite{deWit:1979dzm,deWit:1980lyi,deWit:1984wbb,deWit:1984rvr,Bergshoeff:1980is}. We summarise some of its features that are relevant
for this paper.
In the Poincar\'e frame, after eliminating
the auxiliary fields, the fields are
the spacetime metric $g_{\mu\nu}$, the gravitini $\psi_\mu^i$, the gauge fields $W_\mu^I$, the gaugini $\Omega_i^I$ and the scalar fields $X^I$.  They transform as follows under $Q$-supersymmetry (dropping higher-order
fermionic terms),
\be
\begin{aligned}
\delta_Q g_{\mu\nu} &= \bar{\epsilon}^i \gamma_{(\mu} \psi_{\nu) i} + h.c. \;,  \\
\delta_Q \psi_ {\mu}^i &= {\cal D}_{\mu} \epsilon^i - \tfrac{1}{16}  T_{a b}^{-} \gamma^{ab} \, \gamma_{\mu} \,
 \varepsilon^{ij}\epsilon_j \;, \\
 \delta_Q W^I_{\mu} &= \tfrac12 \varepsilon^{ij} \bar \epsilon_i \left( \gamma_{\mu} \Omega^I_j + 2 \psi_{\mu j}  X^I \right)  + h.c. \;,  \\
\delta_Q \Omega^{Ii} &= \slashed{\cal{D}} \bar{X}^I \, \epsilon^i + \tfrac14 \gamma^{ab} \mathcal{F}_{ab}^{I+} \, \varepsilon^{ij} \epsilon_j \;, \\
\delta_Q X^I &= \tfrac12 \, \bar \epsilon^i \, \Omega^I_i \;,
\end{aligned}
\ee
where the covariant derivatives are given by 
\be
\begin{aligned}
{\cal D}_{\mu} \bar{X}^I &= \left( \partial_{\mu} + i a_{\mu} \right)\bar{X}^I \;, \\
{\cal D}_{\mu} \epsilon^i &= \left( \partial_{\mu} + \tfrac14 \omega_{\mu}{}^{ab} \gamma_{ab} - \tfrac{i}{2} a_{\mu} \right) \epsilon^i  \;,
\end{aligned}
\ee
and we have the composite quantities
\be
\label{comp_fields}
\begin{aligned}
T_{ab}^{-} &= 2 \, \frac{N_{IJ}  \, {\bar X}^J  }{ N_{KL} \,  {\bar X}^K {\bar X}^L } \, F_{ab}^{I-}  \;, \\
\mathcal{F}_{ab}^{I+}&= F_{ab}^{I+} - \tfrac12 X^I \, T_{ab}^+  \;, \\
a_{\mu} &= - \tfrac12  \left( F_I \partial_{\mu} {\bar X}^I - {\bar X}^I \partial_{\mu} F_I + {\rm c.c.} \right) \;.
\end{aligned}
\ee
Here $F_I = \partial F(X) / \partial X^I$, where $F(X)$ denotes the prepotential of the model.
The scalar fields $X^I$ satisfy the Einstein frame constraint
\be
N_{IJ}X^I\bar{X}^J=-1 \;,
\label{einsfra}
\ee
where
\begin{equation}
N_{IJ} = - i \left( F_{IJ} - \bar F_{IJ} \right) \;\;\;,\;\;\; F_{IJ} = \frac{\partial^2 F(X)}{\partial X^I \partial X^J} \;.
\end{equation}
The gaugini $\Omega^I_i$ are constrained by the S-supersymmetry gauge fixing condition
\begin{equation}
{\bar X}^I \, N_{IJ} \, \Omega^J_i = 0 \;.
\end{equation}

Now we focus on the model
$F(X) = -i X^0 X^1$, for which $N_{00} = N_{11} = 0, \, N_{01} = N_{10} = -2$, as well as
\begin{equation}
\Omega_i^1 = - \bar z \, \Omega^0_i \;,
\label{ssusycons}
\end{equation}
where $z = X^1/ X^0$. The Einstein frame constraint \eqref{einsfra} becomes
\begin{equation}
2 \, |X ^0|^2 \, (z + \bar z) = 1 \;.
\end{equation}

Next, we linearise this theory around a flat spacetime background with metric $\eta_{\mu \nu}$ and constant scalar fields $\langle X^I \rangle $,
\begin{eqnarray}
g_{\mu \nu} &=& \eta_{\mu \nu} + h_{\mu \nu} \;, \nonumber\\
X^I &=& \langle X^I \rangle + \delta X^I \;.
\end{eqnarray}
In order to avoid cluttering of notation, we will denote the fluctuations $\delta X^I$ simply by
$X^I$. The linearised spin connection $\omega_{\mu ab}$ and the linearised Riemann tensor 
$R_{\mu\nu ab}= \partial_{\mu} \omega_{\nu ab} - \partial_{\nu} \omega_{\mu ab}$ are given by
(droppping pure gauge terms in $\omega_{\mu ab}$)
\be
\begin{aligned}
\omega_{\mu ab}&=-\partial_{[a}h_{b]\mu} \;, \\
R_{\mu\nu\alpha\beta}&
=-2\partial_{[\alpha}\partial_{[\mu}h_{\nu]\beta]} \;.
\end{aligned}
\ee
The Q-supersymmetry parameter $\epsilon^i$ splits into a local part and a global (rigid) part.
The S-supersymmetry constraint \eqref{ssusycons} results in
\be
\partial_{\mu} \left( \Omega^{1}_i +\langle \bar z \rangle\Omega^{0}_i \right)=0 \;,
\ee
which, upon contraction with ${\bar \epsilon}^i$, equals the supersymmetry variation of
\begin{equation}
 \partial_{\mu} \left( X^1 + \langle {\bar z} \rangle \, X^0  \right)=0 \;.
 \label{aconnvan}
\end{equation}
The Einstein frame constraint \eqref{einsfra} reduces to
\be
X^0\langle\bar{X}^1\rangle+X^1\langle\bar{X}^0\rangle
+\langle X^0\rangle\bar{X}^1+\langle X^1\rangle\bar{X}^0=\frac{1}{2}  \;.
\ee
Using this, the connection $a_{\mu}$ in \eqref{comp_fields} becomes
\be
a_\mu= - 2i  \langle\bar{X}^0\rangle   \left( \partial_\mu X^1+\langle\bar{z} \rangle\partial_\mu X^0 \right) \;,
\ee
which vanishes by virtue of \eqref{aconnvan}. Hence, at the linearised level,
we have
\begin{equation}
a_{\mu} = 0 \;.
\end{equation}
For the other composite fields in \eqref{comp_fields} we obtain, at the linearised level,
\be
\begin{aligned}
T_{\mu\nu}^-& = \frac{1}{\langle\bar{X}^0\rangle}\left[ F_{\mu\nu}^{-0}+\frac{F_{\mu\nu}^{-1}}{\langle\bar{z}\rangle}\right] \;, \\
\mathcal{F}_{\mu\nu}^{0+}& = \frac{1}{2}\left[F_{\mu\nu}^{0+}-\frac{F_{\mu\nu}^{+1}}{\langle z\rangle} \right]  \;,
\\
\mathcal{F}_{\mu\nu}^{1+}& = \frac{1}{2}\left[F_{\mu\nu}^{1+}-\langle z \rangle 
F_{\mu\nu}^{+0} \right]  \;.
\end{aligned}
\ee


\section{Equations of motion}
In this appendix we summarise all the equations of motion of the fields involved, both on the gauge side and on the gravity side. Since we work in the linearised approximation, we note that the equations for the fluctuations of different fields decouple. We also describe the gauge fixing of the non-supersymmetryc field ${\tilde A}_\mu$.  
\subsection{(Super) Yang-Mills}
The equations of motion and Bianchi identities for the fields of the $\N=2$ super Yang-Mills multiplet, in the absence of sources, are
\be
\label{LHS_eom}
\begin{aligned}
\partial_\mu F^{\mu\nu a}=\partial_\mu (^* F)^{\mu\nu a}&=0 \;, \\
\slashed{\partial}\lambda_i^a&=0 \;,\\
\square\sigma^a&=0 \;,
\end{aligned} 
\ee 
and similarlty for the 
source-free $\N=0$ Yang-Mills gauge field,
\be
\label{RHS_eom}
\partial_\mu\tilde{F}^{\mu\nu\tilde{a}}=\partial_\mu(^* \tilde{F})^{\mu\nu\tilde{a}}=0 \:.
\ee
We will find that it is useful to work in a Lorenz like gauge
\be
\label{lorenz_like_gauge}
\partial_\mu\tilde{A}^{\mu\tilde{a}}=0 \;.
\ee
In this gauge the associated equation of motion reduces to 
\be 
\label{eom_tilde_A}
\square \tilde{A}_\mu^{\tilde{a}}=0.
\ee
\subsection{Supergravity}
\subsubsection{Fermionic fields}
In the linearised theory, the gaugini satisfy the Dirac equation
\be
\label{gaugino_eom}
\slashed{\partial}\Omega^{Ii}=0 \;,
\ee
and the gravitini satisfy
\be
\label{gravitino_eom_1}
\gamma^{\mu\nu\rho}\partial_\nu\psi_\rho^i=0, 
\ee
which can be brought into the equivalent form
\be
\label{gravitino_eom_2}
\gamma^\mu[\partial_\mu\psi_\nu^i-\partial_\nu\psi_\mu^i]=\gamma^\mu\psi_{\mu\nu}^i=0. 
\ee
Applying $\partial_{\rho}$ to this and antisymmetrizing in $\rho \nu$ results in 
\be 
\slashed{\partial} \psi_{\rho \nu}^i = 0 \;.
\ee
Finally, by contracting \eqref{gravitino_eom_2}
with $\gamma^\rho\partial_\rho$, we obtain yet another form,
\be
\label{gravitino_eom_3}
\partial^\mu\psi_{\mu\nu}^i=0.  
\ee
\subsubsection{Bosonic fields}
The scalar fluctuations will satisfy the wave equation
\be
\square X^I=0, 
\ee
and the gauge fields will decouple to individually satisfy the Maxwell equation and Bianchi identity
\be
\partial^\mu F_{\mu\nu}^I=\partial^\mu(^*F)_{\mu\nu}^I=0. 
\ee
Finally, in the weak field limit, Einstein's equations reduce to
\be
R_{\mu\nu}=0. 
\label{rmn}
\ee

\section{Dictionary derivation} \label{dderiva}
In this section we describe how the double copy dictionary in \eqref{the_big_dictionary} is obtained. We will use the following simplified notation throughout this appendix,
\be
\varphi^a\star\phi_{a\bar{a}}\star\tilde{\varphi}^{\bar{a}}\equiv
\varphi\star\tilde{\varphi} \;.
\ee 
This is motivated by the fact that the transformation properties of the spectator scalar do not contribute to the derivation of the dictionary.  We work in the Lorentz like gauge $\partial_\mu\tilde{A}^{\mu} =0$.
Additionally, we will make extensive use of the following property of the convolution
\be
\label{non_Leibniz}
\partial_\mu(f\star g)=(\partial_\mu f)\star g=f\star(\partial_\mu g) \;.
\ee
In \autoref{table_on_shell_states}, we present the on-shell tensoring of helicity states. Motivated by this, we begin with the ansatz
\be 
\label{app_ansatz}
a\psi_{\mu\nu}^i+2b\gamma_{[\nu}\partial_{\mu]}\Omega^{0i}
\equiv \varepsilon^{ij}\lambda_j 
\star\tilde{F}_{\mu\nu}  \;,
\ee
with $a,b \in \mathbb{C}$ complex constants, carrying the appropriate chiral weights under $U(1)$ (we note that the chiral weights of the gravitini and the gaugini differ by $\pm 1$, depending on convention). We now contract the above with $\gamma^\mu$ and, making use of the gravitini equation of motion \eqref{gravitino_eom_2}, we get
\be
2b\gamma^\mu\gamma_{[\nu}\partial_{\mu]}\Omega^{0i}=\varepsilon^{ij}\gamma^\mu\lambda_j 
\star(\partial_\mu\tilde{A}_\nu-\partial_\nu\tilde{A}_\mu) \;.
\ee
Now we use the Clifford algebra relation \eqref{gam-decom} together with the equation of motion for the gaugini \eqref{gaugino_eom} on the left hand side (LHS); on the right hand side (RHS) we employ the property \eqref{non_Leibniz}, together with the equation of motion for $\lambda_j$ from \eqref{LHS_eom}, to simplify the expression to
\be 
\label{app_gaugino_dictionary}
2b\partial_\mu\Omega^{0i}=\varepsilon^{ij}\gamma^\rho\lambda_j \star\partial_\mu\tilde{A}_\rho \;.
\ee  
Thus, we have obtained the dictionary entry for the gaugini. The next step is to check whether this double copy expression satisfies the equation of motion for the gaugini \eqref{gaugino_eom}. To verify this, we contract 
\eqref{app_gaugino_dictionary}
with $\gamma^\mu$ to get
\be 
\begin{aligned}
0=2b\gamma^\mu\partial_\mu\Omega^{0i}&=\varepsilon^{ij}\gamma^\mu\gamma^\rho\lambda_j \star\partial_\mu\tilde{A}_\rho\\
&=\varepsilon^{ij}\gamma^\rho\slashed{\partial}\lambda_j \star\tilde{A}_\rho
+2\varepsilon^{ij}\lambda_j\star\partial^\mu\tilde{A}_\mu \;,
\end{aligned} 
\ee  
where to get to the second line we made use of the Clifford algebra relation \eqref{gam-decom} and the convolution property \eqref{non_Leibniz}. The first term now vanishes by \eqref{LHS_eom}, while the second vanishes because $\tilde{A}_\mu$ is taken to satisfy the Lorenz like gauge \eqref{lorenz_like_gauge}. Finally, the LHS of \eqref{app_gaugino_dictionary} must vanish when we impose the equation of motion for $\tilde{A}_\mu$. Working with the form $\square\tilde{A}_\mu=0$, we see that this is indeed the case, since
\be
\square\Omega^I_i=0
\ee
in the linearised theory. Next, we plug in the dictionary for the gaugini into \eqref{app_ansatz} to read off the dictionary for the gravitini,
\be
a\psi_{\mu\nu}^i=\varepsilon^{ij}
[2\partial_{[\mu}\lambda_j\star\tilde{A}_{\nu]}-\gamma_{[\nu}\gamma^\rho\lambda_j \star\partial_{\mu]}\tilde{A}_\rho] \;.
\ee
We can make use of the Clifford algebra relation to rewrite this in the simpler form
\be
 a\psi_{\mu\nu}^i=\varepsilon^{ij} \gamma^\rho\gamma_{[\nu}\lambda_j\star\partial_{\mu]}\tilde{A}_\rho \;.
\ee 
As before, we now proceed to checking whether this dictionary is compatible with the equation of motion 
\eqref{gravitino_eom_2}. Contracting the expression above with $\gamma^\mu$ and using $\gamma^b \gamma_a \gamma_b = 
- 2 \gamma_a$ we obtain
\be
 a\gamma^\mu\psi_{\mu\nu}^i=\frac{1}{2}\varepsilon^{ij} \left( 2 \gamma_{\nu} \lambda_j\star\partial^{\mu}
 \tilde{A}_{\mu} + \gamma^{\rho} \gamma_{\nu} \slashed{\partial} \lambda_j \star \tilde{A}_{\rho} \right) \;,
 \ee
where we made use of the convolution property \eqref{non_Leibniz}. This vanishes by virtue of 
\eqref{LHS_eom} and because $\tilde{A}_\mu$ is taken to satisfy the Lorentz like gauge \eqref{lorenz_like_gauge}.
Moreover, it is easy to see from the above that putting the Yang-Mills fields $\lambda_i$ and $\tilde{A}_\mu$ on-shell exactly corresponds to putting the gravitini on-shell. 

We now proceed to derive the dictionary for the field strengths of the supergravity gauge fields $W_{\mu}^I$. 
It is convenient to 
first work out double copy expressions for the combinations $T_{\mu\nu}^-$ and $\mathcal{F}_{\mu\nu}^{0+}$, given in \eqref{combs_field_strengths}, that appear in the linearized supergravity transformation rules \eqref{all_sugra_SUSY}.
To derive $T_{\mu\nu}^-$, we will make use of the supersymmetry variation of the gravitini field strength,
by comparing terms of like chirality or, equivalently, of the same helicity. Using \eqref{all_sugra_SUSY}, we obtain
\be
\label{T_minus_sugra}
a\delta_Q\psi_{\mu\nu}^i(T)=-\frac{1}{8}\partial_{[\mu}(aT_{\alpha\beta}^-)\gamma^{\alpha\beta} \gamma_{\nu]}\varepsilon^{ij}\epsilon_j \;.
\ee 
Here, the notation $\delta_Q\psi_{\mu\nu}^i(T)$ means that we are only considering terms in the variation 
that are proportional to $T$. Similarly, when writing $\delta_Q\lambda_j(\sigma)$ below, we are only
retaining terms proportional to $\sigma$.
This has to match with the supersymmetry variation on the super Yang-Mills side,
\be 
\begin{aligned}
a\delta_Q\psi_{\mu\nu}^i(T)
&=\varepsilon^{ij} \gamma^\rho\gamma_{[\nu}\left(\delta_Q\lambda_j(\sigma)\right) \star\partial_{\mu]}\tilde{A}_\rho\\
&=\gamma^\rho\gamma_{[\nu}\gamma^\alpha\partial_\alpha\sigma \star\partial_{\mu]}\tilde{A}_\rho\varepsilon^{ij}\epsilon_j\\
&=-\gamma^\rho\gamma^\alpha\partial_\alpha\sigma
\star\partial_{[\mu}\tilde{A}_\rho\gamma_{\nu]}\varepsilon^{ij}\epsilon_j\\
&=-\gamma^\rho\gamma^\alpha\partial_{[\mu}(\sigma
\star\partial_\alpha\tilde{A}_\rho)\gamma_{\nu]}\varepsilon^{ij}\epsilon_j \;.
\end{aligned}
\ee
Here we got to the second line via \eqref{rigid-transf}, to the third line via the Clifford algebra relations, and the final expression is obtained through the convolution property \eqref{non_Leibniz}. 
Using the relation \eqref{gam-decom} once more as well as the 
Lorenz gauge condition $\partial^\rho\tilde{A}_\rho=0$, we are left with
\be
\label{T_minus_YM}
\begin{aligned}
a\delta_Q\psi_{\mu\nu}^i(T)&=
\frac{1}{2}\gamma^{\rho\alpha}\partial_{[\mu}(\sigma
\star\tilde{F}_{\rho\alpha})\gamma_{\nu]}\varepsilon^{ij}\epsilon_j\\
&=\frac{1}{2}\gamma^{\alpha\beta}\partial_{[\mu}(\sigma
\star\tilde{F}_{\alpha\beta}^-)\gamma_{\nu]}\varepsilon^{ij}\epsilon_j \;,
\end{aligned} 
\ee 
where in the last line we have projected out the self-dual part, in light of \eqref{appA_gammas}. Finally, comparing \eqref{T_minus_sugra} and \eqref{T_minus_YM}, we read off
\be
aT_{\mu\nu}^-=-4\sigma\star\tilde{F}_{\mu\nu}^- \;.
\ee
The other composite quantity, $\mathcal{F}_{\mu\nu}^{0+}$, is derived analogously from the supersymmetry variation of the gaugini. Using \eqref{all_sugra_SUSY}, we have
\be
\label{calf_from_sugra}
2b\delta_Q\partial_\mu\Omega^{0i}(\mathcal{F})=\frac{1}{2}\gamma^{\alpha\beta}\partial_\mu(b\mathcal{F}_{\alpha\beta}^{0+}) \varepsilon^{ij}\epsilon_j \;.
\ee
We match this with the supersymmetry variation on the super Yang-Mills side,
\be
\begin{aligned}
2b\delta_Q\partial_\mu\Omega^{0i}(\mathcal{F})&=
\varepsilon^{ij}\gamma^\rho\delta_Q\lambda_j(\sigma)\star\partial_\mu\tilde{A}_\rho\\
&=\gamma^\rho\gamma^\alpha\partial_\alpha\sigma\star\partial_\mu\tilde{A}_\rho
\varepsilon^{ij}\epsilon_j\\
&=\gamma^\rho\gamma^\alpha\partial_\mu(\sigma\star\partial_\alpha\tilde{A}_\rho)
\varepsilon^{ij}\epsilon_j \;,
\end{aligned} 
\ee
where we used \eqref{rigid-transf} to get to the second line and the non-Leibniz behaviour of the convolution \eqref{non_Leibniz} to get to the third line. We again use 
the relation \eqref{gam-decom}
and the Lorenz gauge for $\tilde{A}_\rho$ to write
\be
\label{calf_from_SYM}
\begin{aligned}
2b\delta_Q\partial_\mu\Omega^{0i}(\mathcal{F})&=-\frac{1}{2}\gamma^{\rho\alpha}\partial_\mu(\sigma\star\tilde{F}_{\rho\alpha})
\varepsilon^{ij}\epsilon_j\\ 
&=-\frac{1}{2}\gamma^{\alpha\beta}\partial_\mu(\sigma\star\tilde{F}_{\alpha\beta}^+)
\varepsilon^{ij}\epsilon_j \;,
\end{aligned}
\ee
where we made use of the right-handed version of \eqref{appA_gammas}. Then, comparing \eqref{calf_from_sugra} and \eqref{calf_from_SYM}, we get
\be 
b\mathcal{F}_{\mu\nu}^{0+}=-\sigma\star\tilde{F}_{\mu\nu}^+ \;.
\ee  
We now recall, from \eqref{combs_field_strengths}, that
\be
\begin{aligned}
T_{\mu\nu}^-&=\frac{1}{\langle\bar{X}^0\rangle}\left[ F_{\mu\nu}^{0-}+\frac{F_{\mu\nu}^{1-}}{\langle\bar{z}\rangle}\right]   \;, \\
\mathcal{F}_{\mu\nu}^{0+}&=\frac{1}{2}\left[F_{\mu\nu}^{0+}-\frac{F_{\mu\nu}^{1+}}{\langle z\rangle} \right]  \;,
\end{aligned} 
\ee 
from which we extract
\be 
\begin{aligned}
F_{\mu\nu}^0&=-\sigma\star\left[\frac{2\langle\bar{X}^0\rangle}{a}\tilde{F}_{\mu\nu}^- 
+\frac{1}{b}\tilde{F}_{\mu\nu}^+\right]+h.c. \;,\\
F_{\mu\nu}^1&=-\sigma\star\left[\frac{2\langle\bar{X}^1\rangle}{a}\tilde{F}_{\mu\nu}^-
-\frac{\langle z\rangle}{b}\tilde{F}_{\mu\nu}^+\right]+h.c. \;.
\end{aligned}
\ee
Given that $\tilde{F}_{\mu\nu}^\pm=\frac{1}{2}(\tilde{F}_{\mu\nu}\pm (*\tilde{F})_{\mu\nu})$, it is now easy to see that the equations of motion and Bianchi identities for the 
field strengths $F^I_{\mu \nu}$
are in direct correspondence with those of the Yang-Mills side. One can also check that the LHS and RHS of the above expressions transform identically under supersymmetry. 

We continue with the derivation of the dictionary for the supergravity scalar $X^0$. We make use of the supersymmetry transformation of the gaugini \eqref{all_sugra_SUSY} to write
\be
\label{app_scalar_from_sugra}
2b\delta_Q\partial_\mu\Omega^{0i}(X)=2b\gamma^\rho\partial_\mu(\partial_\rho\bar{X}^0)\epsilon^i \;.
\ee   
We will compare this with the supersymmetry variation on the super Yang-Mills side,
\be
\begin{aligned}
2b\delta_Q\partial_\mu\Omega^{0i}(X)&=
\varepsilon^{ij}\gamma^\rho\delta_Q\lambda_j(F)\star\partial_\mu\tilde{A}_\rho\\
&=-\frac{1}{4}\gamma_\rho\gamma^{\alpha\beta}F_{\alpha\beta}^- \star\partial_\mu\tilde{A}^\rho\epsilon^i \;,
\end{aligned} 
\ee
where we plugged in \eqref{rigid-transf} and used $\varepsilon^{ij} \varepsilon_{jk} = - \delta^i_k$. We now make use of the relation \eqref{app_anti_self_contract}
to rewrite
\be
\label{app_scalar_from_SYM}
\begin{aligned}
2b\delta_Q\partial_\mu\Omega^{0i}(X)&=\gamma^\alpha F_{\alpha\rho}^-\star\partial_\mu\tilde{A}^\rho\epsilon^i\\
&=\gamma^\rho\partial_\mu(F_{\rho\nu}^-\star\tilde{A}^\nu)\epsilon^i \;,
\end{aligned} 
\ee
where we renamed dummy variables and made use of the convolution property \eqref{non_Leibniz}. Then, comparing \eqref{app_scalar_from_sugra} and \eqref{app_scalar_from_SYM}, we read off the dictionary for the scalar
\be 
\label{app_dict_scalar}
b\partial_\mu\bar{X}^0=\frac{1}{2}F_{\mu\rho}^-\star\tilde{A}^\rho \;.
\ee 
As before, we proceed by checking the equations of motion. This is done most easily by contracting the expression above with $\partial^\mu$,
\be
b\square\bar{X}^0=\frac{1}{2}\partial^\mu F_{\mu\rho}^-\star\tilde{A}^\rho \;.
\ee  
Thus we see that the equation of motion for the scalar field follows from the equation of motion and Bianchi identity for the super Yang-Mills gauge field, and vice-versa. Similarly, imposing the equations of motion for
${\tilde A}_{\mu}$ implies the equation of motion for $X^0$.
Additionally, one can show that both sides of the above equation transform identically under supersymmetry. Also, acting with $\partial_\nu$ on \eqref{app_dict_scalar}  and anti-symmetrising in $\mu \nu$, one obtains the relation
\be 
\label{app_useful}
0=F_{\rho[\mu}^-\star\partial_{\nu]}\tilde{A}^\rho \;.
\ee
This relation is satisfied by virtue of the equations of motion and the Lorentz like gauge for ${\tilde A}_{\mu}$.

The final step is to derive the dictionary for the Riemann tensor. We recall that
\be 
R_{\mu\nu\alpha\beta}=-2\partial_{[\alpha}\partial_{[\mu}h_{\nu]\beta]} \;,
\ee
and we make use of the supersymmetry transformation of the gravitino \eqref{all_sugra_SUSY}, 
\be 
\label{app_riemann_sugra}
a\delta_Q\psi_{\mu\nu}^i(R)=\frac{a}{4}\gamma^{\alpha\beta}R_{\mu\nu\alpha\beta}^-\epsilon^i \;,
\ee
where the anti self-dual part is taken over $\alpha\beta$. This is compared with the supersymmetry variation on the super Yang-Mills side \eqref{rigid-transf},
\be  
\begin{aligned}
a\delta_Q\psi_{\mu\nu}^i(R)&=\varepsilon^{ij} \gamma^\rho\gamma_{[\nu}\left(\delta_Q\lambda_j(F)\right) \star\partial_{\mu]}\tilde{A}_\rho\\
&=-\frac{1}{4}\gamma^\rho\gamma_{[\nu}\gamma^{\alpha\beta}F_{\alpha\beta}^- \star\partial_{\mu]}
\tilde{A}_\rho\epsilon^i \;.
\end{aligned}
\ee
We now make use of \eqref{app_anti_self_contract} to simplify the above to
\be
a\delta_Q\psi_{\mu\nu}^i(R)=\gamma^\rho\gamma^\alpha F_{\alpha[\nu}^-
\star\partial_{\mu]}\tilde{A}_\rho\epsilon^i \;.
\ee
At this stage we again use $\gamma^\alpha\gamma^\beta=\eta^{\alpha\beta}+\gamma^{\alpha\beta}$, and, in light of \eqref{app_useful}, we are left with
\be
a\delta_Q\psi_{\mu\nu}^i(R)=\gamma^{\rho\alpha} F_{\alpha[\nu}^-
\star\partial_{\mu]}\tilde{A}_\rho\epsilon^i \;.
\ee
We now recall the definition of the anti self-dual tensor from \eqref{appA_dual_tensors}, and making judicious use of the gamma matrix identities in \eqref{appA_gammas}, together with the equations of motion for the super Yang-Mills gauge field, we derive
\be
\label{app_riemann_squaring}
a\delta_Q\psi_{\mu\nu}^i(R)=-\frac{1}{8}\gamma^{\alpha\beta} \left[F_{\mu\nu}\star \tilde{F}_{\alpha\beta}^{-} +F_{\alpha\beta}^-\star \tilde{F}_{\mu\nu}
-4\eta_{[\alpha[\mu}\partial_{\nu]}\partial_{\beta]}^-A^\rho\star\tilde{A}_\rho\right] \;.
\ee   
Finally, we compare \eqref{app_riemann_sugra} and \eqref{app_riemann_squaring}, and read off
\be
aR_{\mu\nu\alpha\beta}^-=
-\frac{1}{2}\left[F_{\mu\nu} \star \tilde{F}_{\alpha\beta}^- +F_{\alpha\beta}^-\star \tilde{F}_{\mu\nu}
-4\eta_{[\alpha[\mu}\partial_{\nu]}\partial_{\beta]}^-A^\rho\star\tilde{A}_\rho\right] \;.
\ee
This can also be written as
\be
aR_{\mu\nu\alpha\beta}^-=
2\left[F_{[\alpha [\mu} \star \tilde{F}_{\nu ] \beta]} 
+\eta_{[\alpha[\mu}\partial_{\nu]}\partial_{\beta]}A^\rho\star\tilde{A}_\rho\right]^- \;,
\ee
where the anti self-dual part is taken over the $\alpha\beta$ indices. We can then extract the double copy relation for the Riemann tensor,
\be
R_{\mu\nu\alpha\beta}=-\frac{1}{2a}\left[F_{\mu\nu}\star \tilde{F}_{\alpha\beta}^- +F_{\alpha\beta}^-\star \tilde{F}_{\mu\nu}
-4\eta_{[\alpha[\mu}\partial_{\nu]}\partial_{\beta]}^-A^\rho\star\tilde{A}_\rho\right]+h.c. \:.
\ee
Now we check that the Ricci tensor constructed from $R_{\mu\nu\alpha\beta}$ vanishes, hence that
Einstein's equations \eqref{rmn} are satisfied.
One can show that this holds as a consequence of equations of motion and Bianchi identities on the field theory side. Moreover, we have checked that the LHS and RHS of the above equation transform identically under supersymmetry. Finally, it is easy to show that the Riemann tensor satisfies the Bianchi identity $R_{\mu[\nu\alpha\beta]}=0$, as required.

\section{Dyonic BPS black hole solutions}\label{App:Linearising Supergravity Solutions}

We consider dyonic BPS black hole solutions in the model based on $F(X) = -i X^0 X^1$.
These solutions are supported by electric charges $(q_0, q_1)$ and by magnetic charges $(p^0, p^1)$ \cite{Ferrara:1996dd}.

The associated line element is of the form
\be
ds^2=-e^{2g}dt^2+e^{-2g}(dr^2+r^2d\Omega_2^2) \;,
\label{lineelbps}
\ee
with $g = g(r)$ determined by
\be 
e^{-2g}=i(\bar{Y}^IF_I (Y) -Y^I \bar{F}_I (\bar Y)) \;.
\ee
Here $F_I(Y)= \partial F(Y)/\partial Y^I$, with
the $Y^I$ defined by \cite{LopesCardoso:2000qm}
\be
Y^I=e^{-g} X^I\bar{k} \;,
\ee
where $k$ denotes the compensating phase introduced in \eqref{BPSstatic}.
The $Y^I$ 
are determined by the attractor equations
\be
\label{attractor_equations}
\begin{aligned}
Y^I-\bar{Y}^I&=i H^I \;, \\
F_I(Y)-\bar{F}_{\bar I} (\bar{Y})&=iH_I \;,
\end{aligned} 
\ee
where the $(H_I, H^I)$ denote harmonic functions 
\be 
H_I=h_I+\frac{q_I}{r},\quad H^I=h^I+\frac{p^I}{r} \;\;\;,\;\;\; I=0,1 \;,
\ee
with integration constants $h_I \in \mathbb{R}, h^I \in \mathbb{R}$ that satisfy the BPS constraint $h^I q_I = h_I p^I$ \cite{Behrndt:1997ny}. We obtain
\be
\begin{aligned}
Y^0&=-\frac{1}{2}\left( H_1 - i H^0 \right)\;\;\;,\;\;\;
Y^1=-\frac{1}{2} \left( H_0 - i H^1 \right)\;\;\;,\;\;\;
e^{-2g}=H_0H_1 + H^0 H^1 \;.
\end{aligned} 
\ee
We impose the asymptotic normalization condition $e^{-2g}|_{r = \infty} = 1$, which results in
\begin{equation}
h_0 \, h_1 + h^0 \, h^1 = 1 \;.
\label{normh01}
\end{equation}
The black hole horizon is at $r=0$.
These black holes are supported by a complex scalar field,
\begin{equation}
z = \frac{Y^1}{Y^0} = \frac{H_0 - i H^1 }{H_1 - i H^0 } \;,
\end{equation}
and by electric-magnetic fields \cite{LopesCardoso:2000qm}
\begin{eqnarray}
F^I_{tr} = - \partial_r \left( e^{2g} \left(Y^I + {\bar Y}^I \right) \right) \;\;\;,\;\;\; F^I_{\theta \phi} = p^I \, \sin \theta \;.
\label{emff}
\end{eqnarray}

We find it convenient to work with an Eddington-Finkelstein type coordinate $u$ defined by
\begin{equation}
du = dt + e^{-2g} \, dr \;.
\label{finkel}
\end{equation}
In coordinates $(u, r, \theta, \phi)$, the line element \eqref{lineelbps} becomes
\begin{equation}
ds^2 = - e^{2g} du^2 + 2 du dr + e^{-2g} r^2 d\Omega_2^2 \;.
\label{lineeddfink}
\end{equation}

Next, we perform a weak field approximation of the solution. This will be used in the main text to obtain
the double copy structure of this solution. 
At large $r$, $e^{-2g}$ is approximated by
\be
e^{-2g}=1+\frac{Q}{r}+\mathcal{O}(r^{-2}) \;,
\ee
where we have introduced the notation
\be
Q=h_0q_1+h_1q_0 + h^0 p^1 + h^1 p^0 \;.
\ee
The line element \eqref{lineeddfink} becomes
\begin{equation}
ds^2 = -  du^2 +  2 du dr + r^2 \, d \Omega_2^2  + \frac{Q}{r} \, du^2  + Q \, r \,  d \Omega_2^2  \;.
\end{equation}
The resulting metric is of 
form $g_{\mu \nu} = \eta_{\mu \nu} + h_{\mu \nu}$, with the background metric $\eta_{\mu\nu}$ and its inverse
$\eta^{\mu \nu}$
given by
\begin{eqnarray}
\label{flat_metric_finkle}
\eta_{\mu \nu} = 
\begin{pmatrix}
-1 & 1 & & \\
1 & 0 & & \\
& & r^2 & 0 \\
& & 0& r^2\, \sin^2 \theta
\end{pmatrix} \;\;\;,\;\;\; 
\eta^{\mu \nu} = 
\begin{pmatrix}
0 & 1 & & \\
1 & 1 & & \\
& & \frac{1}{r^2} & 0 \\
& & 0& \frac{1}{r^2\, \sin^{2} \theta} 
\end{pmatrix} \;,
\end{eqnarray}
and with 
the fluctuation metric $h_{\mu \nu}$ given by
\begin{equation}
h_{\mu \nu} = diag \, (\frac{Q}{r}, 0, Q \, r , Q \, r \, \sin^2 \theta) \;.
\end{equation}
The scalar fields $X^0$ and $z$ get approximated by
\begin{equation}
X^0 \rightarrow \langle X^0 \rangle + X^0 \;\;\;,\;\;\; z \rightarrow \langle z \rangle + z \;,
\end{equation}
where on the right hand side $X^0$ and $z$ denote fluctuating fields. Using \eqref{normh01}, we obtain
\begin{eqnarray}
\langle X^0 \, \bar k \rangle = - \frac12 \, \left( h_1 - i h^0 \right)
\;\;\;,\;\;\; \langle z \rangle = \alpha \;,
\nonumber\\
X^0 \, {\bar k} = - \frac{ \left(q_1  -i p^0 - \tfrac12 Q ( h_1 - i h^0) \right) }{2 r} \;\;\;,\;\;\; z = \frac{\Sigma}{r} \;,
\label{flucX0z}
\end{eqnarray}
where
\begin{equation}
\alpha = \frac{h_0-i h^1 }{h_1- i h^0} \;\;\;,\;\;\; \Sigma = \frac{q_0  -i p^1 - \alpha( q_1 - i p^0) }{h_1 -i h^0} \;.
\end{equation}
Finally, the electric field strengths are approximated by
\begin{equation}
F^0_{ur} =   \frac{Q h_1 - q_1}{ r^2}
\;\;\;,\;\;\; F^1_{ur} =  \frac{Q h_0 - q_0}{r^2} \;,
\label{F01R}
\end{equation}
while the magnetic field strengths are exact and given in \eqref{emff}.

\bibliographystyle{JHEP}

\providecommand{\href}[2]{#2}\begingroup\raggedright\endgroup

\end{document}